\documentclass[aps,prl,twocolumn,groupedaddress]{revtex4}

\usepackage{graphicx}

\def\be{\begin{equation}}
\def\ee{\end{equation}}
\def\ba{\begin{eqnarray}}
\def\ea{\end{eqnarray}}

\begin{document}

\title{Viscous fingering in volatile thin films}

\author{Oded Agam}
\affiliation{The Racah Institute of Physics, The Hebrew University,
Jerusalem, 91904, Israel}


\begin{abstract}

A thin water film on a cleaved mica substrate undergoes a first order phase transition between two values of film thickness. By inducing a finite evaporation rate of the water, the interface between the two phases
develops a fingering instability similar to that observed in the Saffman-Taylor problem. We draw the
connection between the two problems, and construct solutions describing the dynamics of evaporation in this system.
\end{abstract}

\maketitle

The processes of evaporation and dewetting of thin liquid films are becoming
increasingly important in modern technologies e.g. the fabrication of electronic chips, microfluidic
devices, and biosensors. These nonequilibrium processes depend on various details such as the temperature
distribution in the sample, the nature of the interaction between the substrate and the fluid, and the
substrate roughness. Depending on these details, a thin liquid film may, for instance, break into isolated
droplets, evaporate uniformly, or exhibit a fingering instability similar to that observed in the
Saffman-Taylor problem\cite{Saffman58} where an inviscid fluid penetrates a viscous fluid.
This issue naturally generated extensive experimental and theoretical studies -  see
Ref.~\onlinecite{review1} for a recent review of the field. However, due to its complexity, most theoretical studies of the problem are based on numerical computations. In this paper we draw the connection between the evolution of evaporating thin films and the Hele-Shaw type dynamics. This connection allows us to construct analytical solutions for the evolution of evaporation.
Examples for these solutions are depicted in Fig.~1.

The experimental system showing this kind of behavior\cite{Experiment}, consists of a clean mica substrate
covered by a thin water film. Under proper conditions of temperature and pressure, two films of different
thickness (approximately 2 and 12 nm) can coexist. This behavior is a consequence of the competing van der
Waals and polar surface forces between the water and the substrate\cite{Samid98}. A first-order phase
transition between the film heights is induced by changing the vapor pressure.
On the scale of nanometers, gravitational effects are negligible, and if the evaporation rate is sufficiently small one may neglect changes in the film thicknesses due to pressure gradients.
This implies that the dynamics of evaporation, in this system is, effectively, that of the interface between the phases.

We begin by constructing the equations of motion, and from now on refer to the film domain of higher
thickness as a "droplet". Owing to the small thickness of the water film (in both phases), the fluid
dynamics is viscous, and the lubrication approximation with no-slip boundary conditions at the
substrate-liquid interface and zero shear at the liquid-air interface may be employed. Thus, the velocity of the fluid along a profile perpendicular to the substrate, is approximately parabolic. Averaging over this
profile one obtains Darcy's law,
\begin{eqnarray}
{\bf v} =-\frac{ h^2}{3\nu} \nabla P,  \label{eq:darcy}
\end{eqnarray}
\begin{figure}[ptb]
\includegraphics[width=4.5cm]{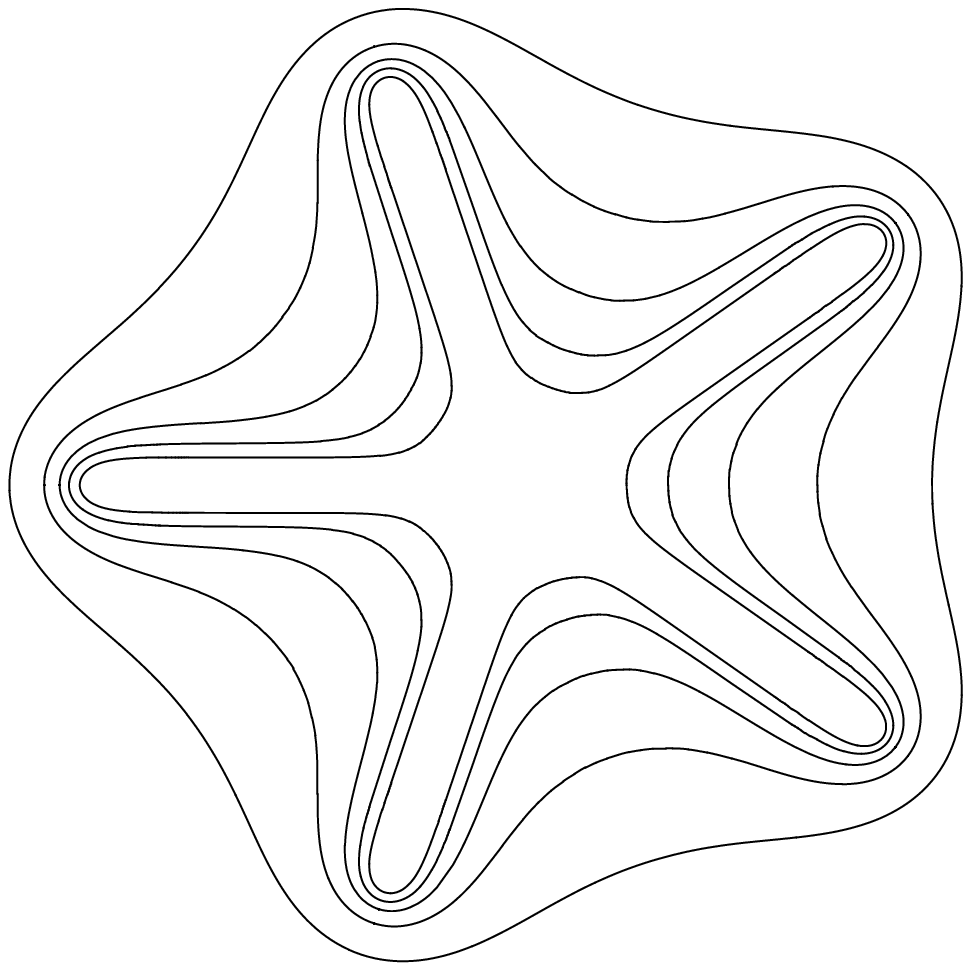}
\includegraphics[width=2.5cm]{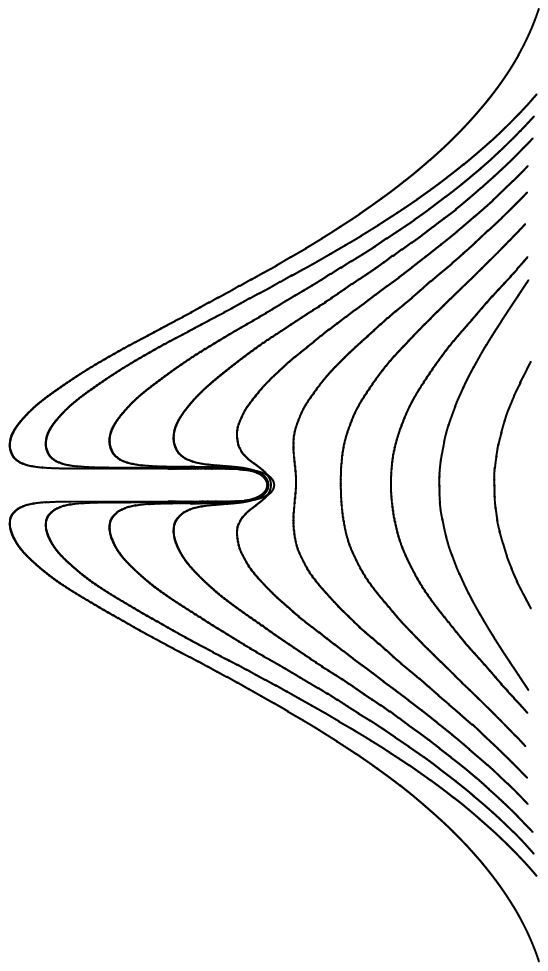}
\label{fig:mineev1} \caption{The evolution of evaporating droplets: Left-hand panel shows snapshots of the
contour of a droplet with fivefold symmetry. Right-hand panel shows a dry patch penetrating, from right to left,
into the liquid domain. In both cases successive contours differ by a fixed time interval.}
\end{figure}
where $h$ is the film height, $\nu$ is its viscosity, $P$ is the pressure, and ${\bf v}$ is the (two-dimensional) velocity field. The dynamics outside the droplet may be ignored since the fluid velocity in
these regions is approximately zero, and evaporation is negligible due to the strong adhesion force between
the substrate and the fluid. Thus evaporation takes place, effectively, only within the droplet where it is
assumed to be uniform. Assuming the film height to be approximately fixed, the variation in the two-dimensional fluid density, $\rho$, may be also neglected and therefore uniform evaporation implies that
\begin{eqnarray}
\rho \nabla \cdot {\bf v} =- \kappa,  \label{eq:evaporation}
\end{eqnarray}
where $\kappa$ is the rate of evaporation (in units of mass per unit time per unit area).
 Combining the above equations one concludes that, inside the droplet domain, ${\cal D}$, the pressure
 satisfies Poisson's equation, while outside this region, it is constant. Without loss of generality we set
 this constant to be zero, and thus
\begin{eqnarray}
 \begin{array}{cc} \nabla^2 P= \beta & \mbox{within~~~} {\cal D}, \\
P= 0 & \mbox{otherwise}, \end{array} \label{eq:Poisson}
\end{eqnarray}
where $\beta=3 \nu \kappa/(h^2 \rho)$. The above equation describes the pressure away from
the droplet's interface. Near that interface, additional degrees of freedom,
such as the film height and the boundary curvature, come into play. The local behavior near the droplet's
edge sets the boundary conditions for the above equations. In the regime where the surface tension
energy associated with the interface can be neglected, the pressure is continuous and therefore the boundary condition for Eq.~(\ref{eq:Poisson}) is
\begin{eqnarray}
 P(z)=0 ~~~\mbox{for}~~ z \in \partial {\cal D},  \label{eq:boundary}
\end{eqnarray}
where,  $\partial {\cal D}$ denotes the interface contour. Hereafter $z=x+iy$ will represent
a complex coordinate on the mathematical complex plane where the droplet resides. Eqs.~ (\ref{eq:Poisson})
and (\ref{eq:boundary}) are equivalent to those of the "stamping problem"\cite{Entov92, Gustafsson04}.
The latter is associated with the evolution of droplets in Hele-Shaw cell when the gap between the two
parallel plates changes in time .

Important characteristics of the droplet's shape are provided by the set of the (interior) harmonic
moments:
\begin{eqnarray}
 t_k=\frac{1}{\pi} \int_{z \in {\cal D}(t)} dx dy z^k. \label{eq:tks}
\end{eqnarray}
 where $k$ is a non-negative integer, and the integration is over the droplet's domain (here, for clarity,
 we indicate its explicit time dependence). Following Entov and  \'{E}tingof\cite{Entov92}, we take the time
 derivative of the above equation,
\begin{eqnarray}
 \frac{dt_k}{dt}=\frac{1}{\pi}\oint_{z \in \partial {\cal D}(t)} dl z^k v_n. \label{eq:dttk}
\end{eqnarray}
Here $v_n=-(h^2/3\nu)\partial_n P$ is the interface velocity, where $\partial_n$ is the derivative normal to the interface. Since $P$, vanishes on the boundary, one may add to the integrand a term of the form
$(h^2/3\nu) P \partial_n z^k$. This enables us to use Green's theorem,
\begin{eqnarray}
 \frac{dt_k}{dt}&=&\frac{h^2}{3\pi \nu}\oint_{\partial {\cal D}(t)} dl \left(P\partial_n
 z^k-z^k \partial_n P\right)  \nonumber \\
&=&\frac{h^2}{3\pi \nu} \int_{{\cal D}(t)}dx dy\left(P\nabla^2 z^k- z^k \nabla^2
 P \right).
\end{eqnarray}
However, analyticity implies that $\nabla^2 z^k=0$, and substituting Eq.~(\ref{eq:Poisson}) one immediately
obtains $dt_k/dt=- t_k/\tau$. Thus all $t_k$ decay at the same rate,
\begin{eqnarray}
t_k = t_k(0) e^{-t/\tau}, ~~~\mbox{where} ~~ \tau=\rho/\kappa. \label{eq:tk}
\end{eqnarray}

Knowing the precise time dependence of the harmonic moments allows one to construct exact solutions for the
evolution of the droplet. In particular, the exponential decrease of $t_0$, which is proportional to the droplet's area, implies that the interface velocity of a circular droplet is proportional to its radius, in agreement with the experimental data \cite{LipsonNature}.

It is instructive to consider the evolution of a droplet whose initial shape is an ellipse. The harmonic
moments in this case vanish for all $k>2$, and since Eq.(\ref{eq:tk}) implies no generation of new moments,
the shape of the droplet at any later time remains elliptical. Thus, the pressure takes the form
\begin{figure}[ptb]
\includegraphics[width=4cm]{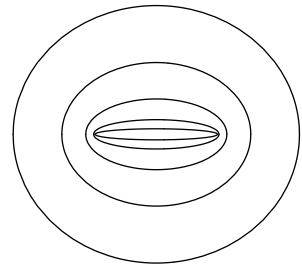}
\label{fig:ellipse} \caption{The evolution of an evaporating elliptical droplet}
\end{figure}
$P=(\beta/4) \mbox{Re} \left[ |z|^2 - \alpha(t) z^2- \gamma(t)\right], \label{eq:p}$
where $\alpha(t)$ and $\gamma(t)$ are time-dependent functions to be determined (linear terms in $x$ and $y$
are omitted since they lead merely to a trivial shift of the droplet).
This form of $P$ satisfies Poisson's equation (\ref{eq:Poisson}), while the boundary condition
(\ref{eq:boundary}) is satisfied on an ellipse,
\begin{eqnarray}
\frac{x^2}{a_+^2(t)}+\frac{y^2}{a_-^2(t)}=1, \label{eq:ellipse}
\end{eqnarray}
provided $|\alpha(t)|< 1$ and $\gamma(t)>0$. Here $a_\pm$ are the semimajor and semiminor  axes
of the ellipse,
\begin{eqnarray}
a_\pm^2 =\frac{ \gamma(t)}{1 \mp \alpha(t)}, \label{eq:ab}
\end{eqnarray}
where without loss of generality we choose $\alpha(t)$ to be real and positive. From Darcy's law
(\ref{eq:darcy}) one can evaluate the velocity of the boundary and deduce that $\partial_t a_\pm =- [1 \mp
\alpha (t)] a_\pm/(2\tau)$. The solution of this equation and (\ref{eq:ab}) yields
\begin{eqnarray}
\alpha(t)= \frac{1}{\sqrt{ 1 + A e^{-2 t/\tau}}}, ~~~~\gamma(t)=  \frac{B e^{-2 t/\tau }}{ \sqrt{1+ A e^{-2
t/\tau}}}, \label{eq:ellipseparameters}
\end{eqnarray}
where $A$  and $B$ are constants determined by the initial shape of the droplet. Substituting
(\ref{eq:ellipseparameters}) into (\ref{eq:ab}) we obtain
\begin{eqnarray}
a_\pm^2(t)=\frac{B e^{-2 t/\tau}}{\sqrt{1+Ae^{-2 t/\tau}}\mp 1}. \label{eq:apm}
\end{eqnarray}
The evolution of an elliptical droplet, as follows from (\ref{eq:ellipse}) and (\ref{eq:apm}), is depicted
in Fig.~2. It shows that the droplet, eventually, contracts to the straight line which connects the ellipse
foci. In more general cases the droplet may contract to a skeleton of line segments called a "mother body"
\cite{Gustafsson04}. In reality, however, surface tension effects neglected here are expected to modify the
evolution at its late stage, and to break the narrow ellipse to a line of small circular droplets.

To describe the evaporation of droplets with more complicated shapes, we shall employ conformal maps from
the interior of the unit circle ($w$ domain) to the interior of the droplet ($z$ domain). A simple example
of such a map is the Joukowski map,
\begin{eqnarray}
z(w)= r(t) w + \frac{u_0(t)}{w-c(t)} + u_1(t) \label{eq:CM}
\end{eqnarray}
with $|c(t)|>1$. The time evolution of the droplet is determined by the time dependence of the coefficients
of the map,  $r(t)$, $c(t)$, $u_0(t)$, and $u_1(t)$ in the above example. This time dependence is dictated
by the exponential decay of the harmonic moments (\ref{eq:tk}). In order to see how, let us introduce the
Schwarz function\cite{Davis74}, $S(z)$, associated with the droplet's boundary in the $z$ plane.  This function is analytic in a strip containing the boundary, and satisfies the condition $\bar{z}=S(z)$ for $z\in
\partial {\cal D}$. Hereafter, an overbar will denote complex conjugation. Expressed in terms of the
conformal map from the interior of the unit circle to the droplet domain, the Schwarz function takes the
form
\begin{eqnarray}
S(z)= \bar{z}\left( \frac{1}{w(z)}\right), \label{SZ}
\end{eqnarray}
where $w(z)$ is the inverse map from the droplet boundary, in the  $z$ plane, to the unit circle in the $w$ plane. This is because, on the unit circle, $\bar{w}=1/w$, and therefore $S(z)=\overline{z(w)}= \bar{z}(\bar{w})
=\bar{z}(1/w)$.

The relation between the Schwarz function and the harmonic moments of the droplet's domain is revealed by
 expressing the integral (\ref{eq:tks}), using Green's theorem, as a contour integral along the droplet's
 boundary,
\begin{eqnarray}
 t_k=\frac{1}{2\pi i} \oint_{\partial {\cal D}(t)} dl S(z) z^k \label{eq:tkSchwarz}
\end{eqnarray}
 Knowing the time dependence of the harmonic moments, Eqs.~ (\ref{SZ}) and (\ref{eq:tkSchwarz}), determine
 the time evolution of the conformal map $z(\omega)$.

Consider, e.g., the Joukowski map (\ref{eq:CM}), where without loss of generality, we choose its
coefficients to be  real. The Schwarz function, in this case, takes the form
\begin{eqnarray}
S(z)= \frac{r(t)}{w(z)} + \frac{u_0(t) w(z)}{1-c(t) w(z)} + u_1(t).  \label{eq:JSF}
\end{eqnarray}
Note that the inverse map $w(z)$, being a solution of quadratic equation, is defined on
two Riemann sheets, thus $S(z)$ is analytic on a Riemann surface of genus zero constructed by gluing
the two patches along the cut, which lies outside the droplet domain. Two poles of $S(z)$ are located inside
${\cal D}$ on the physical sheet, while two other poles are at $z \to \infty$ of both sheets. Thus within
the droplet domain
\begin{eqnarray}
S(z)\sim \sum_{j=1}^2 \frac{\mu_j}{z-q_j}, \label{eq:singular}
\end{eqnarray}
with no other singularities. Performing the contour integral (\ref{eq:tkSchwarz}), by the residue theorem,
we obtain $t_k= \sum_j \mu_j q_j^k$. Now, since all $t_k$ decay at the same rate, $q_j$ should be time
independent, while $\mu_j$ should decay as $\mu_j(t)=\mu_j(0)e^{-t/\tau}$.

To relate the parameters of the Schwarz function, $q_j$ and $\mu_j$, to those of the conformal map
$r$, $u_j$, and $c$, we require both expressions, (\ref{eq:JSF}) and (\ref{eq:singular}), to have the same
poles and residues. This condition leads to the relations
\begin{eqnarray}
q_1&=& z(0), ~~~~~~~~~~ q_2= z(1/c),  \\ ~~~~\mu_1&=&  r
\left. \frac{dz}{dw}\right|_{w=0}, ~~\mu_2= - \frac{u_0}{c^2}
\left. \frac{dz}{dw}\right|_{w=1/c}. \nonumber
\end{eqnarray}
Thus knowing the time dependence of $q_j$ and $\mu_j$, determines
the time dependence of the coefficients of the Joukowski map, and hence the evolution of the droplet.
From the above equations we obtain
\begin{eqnarray}
q_1= 0 =-\frac{u_0}{c}+u_1, ~~~~~ q_2=\frac{r}{c}+ \frac{u_0 c}{1-c^2}+u_1, \nonumber \\
\mu_1(0) e^{- t/\tau}= r \left( r- \frac{u_0}{c^2} \right),~~~~~~~~~~~~~~~ \\
\mu_2(0) e^{- t/\tau}= u_0\left(\frac{u_0}{(c^2-1)^2}-\frac{r}{c^2}\right),~~~~~~~~ \nonumber
\end{eqnarray}
where $q_j$ and $\mu_j(0)$ are determined by the initial shape of the droplet, and without loss of
generality we choose $q_1=0$. The evolution that follows from the above equations, with the initial
conditions $r(0)=1/100$, $u_0(0)=-1/2$, $u_1(0)=-5/12$, and $c(0)=6/5$, is demonstrated in Fig.~3. It
describes a droplet pinching of the puddle. The above equations hold up to the point where the droplets
separate. Beyond this point the two droplets continue to evolve independent of each other. Namely in this
particular case the droplets remain circular while their area decrease exponentially in time.
\begin{figure}[ptb]
\includegraphics[width=4.5cm]{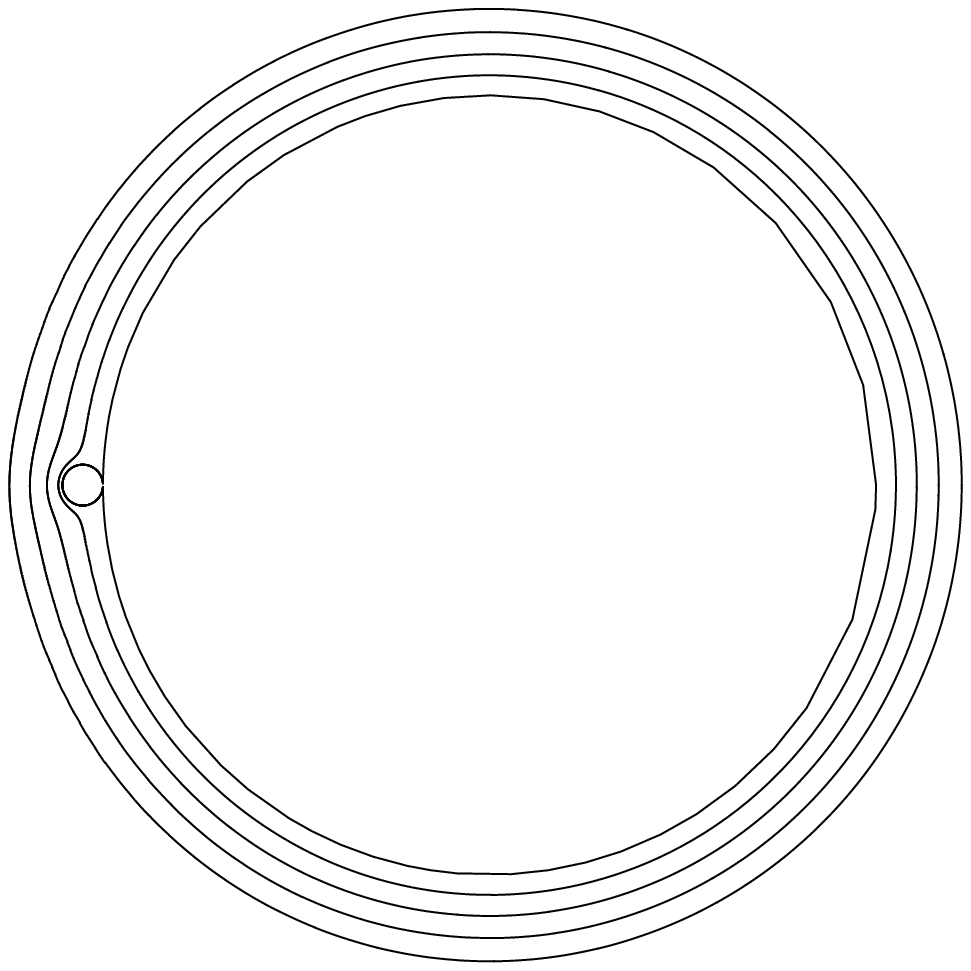}
\includegraphics[width=2cm]{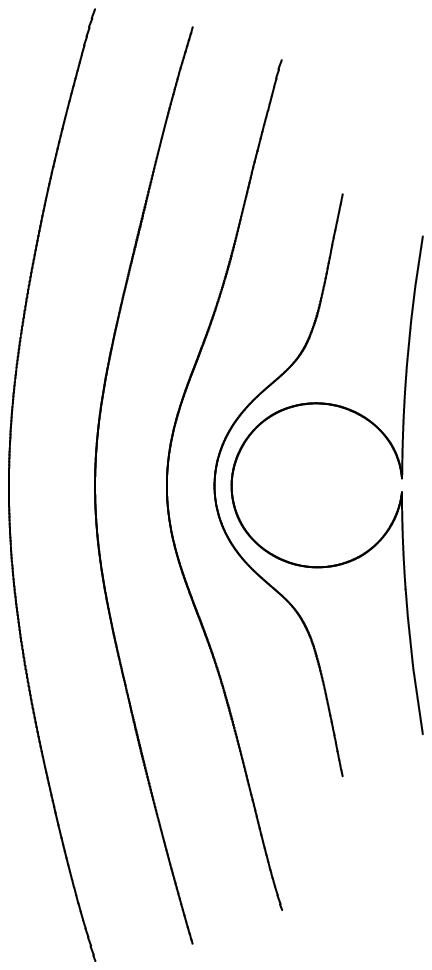}
\label{fig:pinching} \caption{The evolution of droplet pinching due to evaporation. Left-hand
  panel shows the full droplet while the right-hand panel is a
  magnification of the area where the droplet pinches off.}
\end{figure}

Let us now describe the evolution of droplets whose boundaries are described by
conformal maps of the form\cite{Mineev94}
\begin{eqnarray}
z(w)= r w + \sum_j \alpha_j \ln \left( 1- \frac{w}{w_j} \right) \label{eq:MCM}
\end{eqnarray}
with the conditions $|\omega_j|>1$ and  $\sum_j \alpha_j =0$. As in the previous case,
to extract the time dependence of the coefficients, $r$, $\alpha_j$, and $w_j$ we shall identify the
singularities of the Schwarz function, and require them to agree with the time dependence dictated by
(\ref{eq:tk}) and (\ref{eq:tkSchwarz}).

From (\ref{SZ}) and  (\ref{eq:MCM}) it follows that the singular points of $S(z)$ in ${\cal D}$, denoted
henceforth by $q_j$, are the points for which $q_j= z(1/w_j)$. The behavior of the Schwarz function near
these points is given by
\begin{eqnarray}
S\left(q_j+\delta z\right)  = \bar{z} \left(\frac{1}{w\left[z\left( \frac{1}{w_j} \right)+\delta
    z\right]}\right)\simeq \alpha_j
    \ln(\delta z),
\end{eqnarray}
and therefore $S(z)  \sim \sum_j \alpha_j \ln (z-q_j)$. The conditions $|\omega_j|>1$ and  $\sum_j \alpha_j
=0$, ensures that no branch cut crosses the droplet boundary, and from (\ref{eq:tkSchwarz}), we obtain
$t_k=-\sum_j \alpha_j q_j^{k+1}/k$, for $k>0$. Thus, the uniform exponential decay of the harmonic moments
implies that $\alpha_j(t)=\alpha_j(0)e^{-t/\tau}$, while $q_j$ are constants of motion. The equation for
$t_0$, is obtained by deforming the contour of the integral (\ref{eq:tkSchwarz}) around the pole at
infinity. The singular behavior of the Schwarz function in this region is
\begin{eqnarray}
S(z) \to  \bar{z}\left(\frac{r}{z}\right) = \frac{\bar{r}r}{z}-
  \sum_j \bar{\alpha}_j \frac{ r}{\bar{w}_j z}, ~~~~ z\to \infty.
\end{eqnarray}
and therefore the zeroth harmonic moment is  $t_{0} =|r|^2-r
\sum_j \frac{\bar{\alpha}_j}{\bar{w}_j}$. This relation
provides the additional equation needed in order to determine the coefficients of the conformal map
(\ref{eq:MCM}):
\begin{eqnarray}
q_j =  z(w_j^{-1}), ~~~~~~ \alpha_j(0)e^{- t/\tau} = \alpha_j(t), \\
t_{0}(0) e^{- t/\tau} = |r|^2-r \sum_j \frac{ \bar{\alpha}_j(t)}{\bar{w}_j}.~~~~~~  \nonumber
\end{eqnarray}

Two examples of the evolution generated by conformal maps of the type (\ref{eq:MCM})
are depicted in Fig.~1\cite{parameters}. The left-hand panel shows the evolution of a 
droplet contracting to a star-shape form were each one of its arms is similar to 
the Saffman-Taylor finger in a channel
geometry\cite{Saffman59}. In reality these water fingers eventually break into small droplets due to surface tension. Nevertheless, they have been seen in numerical simulations of related problems\cite{numerics}, and
their traces are observed in the experimental data\cite{Experiment}. These studies also
reveal that a rather stable pattern, that commonly appears in these systems, is a dry patch finger split
in the middle by a narrow water "pipe", similar to the result we obtained shown in the right panel of
Fig.~1. The common appearance of these fingers, called  "doublons", in volatile thin films is in a marked
difference with the Saffman-Taylor evolution in a radial geometry, where tips split repeatedly.

It should be noted that Eqs.~ (\ref{eq:Poisson}) and (\ref{eq:boundary}) also have solutions which evolve
into cusp singularities as well as tip splitting, unobserved in the experimental data. An additional
information is needed in order to select the relevant solutions. This information comes from the local
behavior of the fluid near the droplet boundary. There is evidence that Rayleigh instability generates
inhomogeneous distribution of the liquid along the rim of the droplet\cite{Leizerson04}. It is expected that this behavior selects the relevant solutions, such as the finger shown on the right-hand panel of Fig.~1. This
problem is left for further studies.

I thank Paul Wiegmann, Anton Zabrodin, Michael Elbaum, Haim Diamant, Omri Gat, and especially Eldad Bettelheim for useful discussions, and Steve Lipson, Raphi Matthews and Dror Orgad
for reading the manuscript before submission. This work was supported by the United States-Israel Binational Science Foundation (BSF) Grant No.~2004128.


\begin{thebibliography}{99}
\bibitem{parameters} The initial parameters of the doplet shown in the left panel are $w_{\pm 1}=-1.1e^{\pm
    i 2\pi/5}$,$w_{\pm 2}=-1.1e^{\pm i 4\pi/5}$, $w_3=-1.1$, $w_4=1000$, $\alpha_{\pm 1}=3 e^{\pm i
    2\pi/5}$, $\alpha_{\pm 2}=3 e^{\pm i 4\pi/5}$,$\alpha_3=3$, and $r=1$, while those of the right panel
    are $\omega_{\pm 1}=-1.1 e^{\pm i 8\pi/9}$, $\omega_2= 1.3$, $\omega_3=1000$, $\alpha_{\pm 1}= 2 e^{\pm
    i 8\pi/9}$, $\alpha_2=-0.1$, and $r=1$.
\end{thebibliography}
\end{document}